\documentclass{svjour3}
\newcommand{\beq}{\begin{equation}}
\newcommand{\eeq}{\end{equation}}
\newcommand{\beqa}{\begin{eqnarray}}
\newcommand{\eeqa}{\end{eqnarray}}
\newcommand{\Eq}[1]{Eq.\ (\ref{#1})}
\newcommand{\Eqs}[2]{Eqs.\ (\ref{#1}) and (\ref{#2})}
\newcommand{\Eqss}[3]{Eqs.\ (\ref{#1}), (\ref{#2}) and (\ref{#3})}
\newcommand{\Ref}[1]{Ref.\ \cite{#1}}
\newcommand{\Refs}[2]{Refs.\ \cite{#1} and \cite{#2}}

\newcommand{\Section}[1]{Section\ \ref{#1}}

\usepackage{xcolor}  %BORRAR DESPUES DE HACER LOS CAMBIOS!

\journalname{Eur. Phys. J. D}

\begin{document}

\title{
  Simplified calculations of plasma oscillations with
  non-extensive statistics
}

\author{Jos\'e F. Nieves
  \and John D. Verges
}

\institute{
  Laboratory of Theoretical Physics,
  Department of Physics, University of Puerto Rico\\
  17 Ave Universidad Ste 1701, San Juan, Puerto Rico 00925-2537\\
  \email{nieves@ltp.uprrp.edu}
  \email{verges@ltp.uprrp.edu}
}

\date{January 2020}

\maketitle

\begin{abstract}
  We use the exponential parametrization of the nonextensive
  distribution to calculate the dielectric constant in an electron gas
  obeying the nonextensive statistics. As we show, the exponential
  parametrization allows us to make such calculations in a straightforward way,
  bypassing the use of intricate formulas obtained from integral tables
  and/or numerical methods.
  For illustrative purposes, we apply first the method to the calculation of
  the permittivity and the corresponding dispersion relation
  in the ultrarelativistic limit of the electron gas, and verify that it
  reproduces in a simple way the results that had been obtained previously
  by other authors using the standard parametrization of the nonextensive
  distribution. In the same spirit we revisit the calculation of the 
  same quantities for a non-relativistic
  gas, in the high frequency limit, which has been previously carried out,
  first by Lima, Silva and Santos, and subsequently revised by Chen and Li.
  Our own results agree with those obtained by Chen and Li.
  For completeness, we also apply the
  method the low frequency limit in the non-relativistic case,
  which has been previously considered by Dai, Chen and Li
  in the context of the stream plasma instability.
  We discuss some features of the results obtained in each case
  and their interpretation of terms of generalized nonextensive quantities,
  such as the Debye length $\lambda_{D}^{(q)}$, the plasma frequency
  $\omega_{p}^{(q)}$ and the ultra-relativistic frequency
  $\Omega^{(q)}_{e,rel}$. In the limit $q \rightarrow 1$ such quantities
  reduce to their classical value and the classical result of
  the dispersion relations are reproduced.
\end{abstract}

\section{Introduction and Summary}
\label{sec:introduction}

The formalism of the nonextensive
statistics\cite{tsallis,tsallis2,oikonomou,presse,bidollina,footnote1}
has been applied to a variety of physical systems and continues to be
an active field of research. This formalism has been studied in physical
systems where deviations from local thermodynamic
equilibrium occur, for example in the context of heavy ion collisions\cite{dash}
and high energy hadron collisions\cite{smbat}.

Recently, the formalism has been applied in the context of plasma
physics to study the dielectric properties of a
collisionless electron gas, assuming that the underlying
statistics of the background particles obey the nonextensive,
rather than the classical Boltzman
statistics\cite{lima,chenli,liyan,munoz,liu,dai,pain}.
In particular, the nonextensive dielectric permittivity
in the high frequency limit, and the corresponding
dispersion relation, were calculated in \Ref{lima} for the case
of the non-relativistic (NR) electron gas.
That calculation has been revised in \Ref{chenli}, where a different
result was obtained for those same quantities.
In general, the calculations involving the nonextensive
statistics are cumbersome, and usually require the use
of intricate formulas from integral tables. This inconvenience
becomes more pronounced when conflicting results are obtained, such
as those mentioned above.

In the present present work we take another look at
the calculation of the nonextensive dielectric permittivity and dispersion
relation of the collisionless electron gas
using a different method. Our approach consists 
in using an exponential parametrization of the nonextensive
momentum distribution function. Specifically,
as will be discussed in detail in \Section{sec:expparam},
instead of the expression 
\beq
\label{expqdef1}
e_{q}(x) \equiv [1 + \left(1 - q\right)x]^{\frac{1}{1 - q}}\,,
\eeq
that is commonly used\cite{tsallis} for the non-extensive
distribution function, written in terms of generic variable $x$,
we write $e_{q}$ in the form
\beq
\label{expqdef2}
e_{q}(x) = \frac {1}{\Gamma\left(\frac{1}{1 - q}\right)} \int_0^\infty dt\;
t^{\left(\frac{1}{1 - q} - 1\right)} e^{-[1 + (1 - q)x]t}\,.
\eeq
The parameter $q$, which characterizes the non-extensive
distribution function, must be restricted to be
within specific ranges, to be disussed in detail below.
As we will see, this allows us to calculate such quantities
in a straightforward fashion, in particular avoiding the need of making
extensive use of intricate integral table formulas.
Our objective is twofold. On one hand, to verify that
the previously known results are easily reproduced in this way,
and on the other to show how to extend the previous calculations by considering
other limiting cases of practical interest.

To this end, we first consider the case of an ultrarelativistic (ER)
electron plasma, which has been already studied in
\Refs{munoz}{liu}.
Those previous calculations were based on using directly the
standard form of the nonextensive distribution [\Eq{expqdef1}]
provoking either the heavy use of intricate formulas from integral tables
or numerical methods. Here we perform the relevant calculations
using the exponential parametrization method.
As we show, the calculations become straightforward in this manner and
illustrate the effectiveness of the method to treat this kind of
problem. The results we obtain for the dielectric permittivity
and the dispersion relation coincide with those in
\Refs{munoz}{liu}.
In the process we identify the generalized relativistic electron plasma
frequency  $\Omega_{e,rel}^{(q)}$, the temperature $T_{q}$,
and the dielectric permittivity. 
All such physical quantities reduce to their classical counterparts in
the limit  $q \rightarrow 1$, as expected.
This first result illustrates the effectiveness of the method
we propose, and provides confidence on its application.

We then consider two other cases, namely the high and the low frequency limit
of the non-relativistic (NR) plasma. As we have mentioned, the calculation of
the dielectric permittivity in the high frequency limit
has already been considered using the standard parametrization of
the nonextensive statistics distribution in \Refs{lima}{chenli},
with conflicting results. Our own result agrees with the result obtained
in \Ref{chenli} for that same case.  For completeness,
we also consider the NR case in the low frequency limit,
which has been considered before in \Ref{dai} in the context
of the stream instability induced by ions and electrons with
different drift velocities in a dusty plasma. The dielectric permittivity
in that limit can be deduced from the results presented in \Ref{dai} by
adapting and restricting the attention to the electron contribution,
and our own results agree with it. Since the system considered
in \Ref{dai} is a complicated one and the calculation is involved,
this case once again illustrates the effectiveness of the exponential
parametrization method for this kind of calculation. For example,
we show that the next-to-leading order contribution
to the dielectric permittivity (in powers of $\omega/k$),
which is not considered in \Ref{dai}, is obtained equally simply.
We identify a temperature parameter $T_{q}$, 
in analogy with the high frequency limit case, as well as a
parameter $\lambda_{D}^{q}$ that is
analogous to the Debye length of a classical plasma,
both of which depend on the parameter $q$ that characterizes the nonextensive
distribution.  With such identifications, the
resulting formula for the dielectric permittivity in the low frequency limit
can be expressed in the same form as the classical case,
but with the paramemeters $T_{q}$ and $\lambda_{D}^{q}$ replacing
the temperature and Debye length, respectively. In the limit
$q \rightarrow 1$, both $T_{q}$ and $\lambda_{D}^{q}$ reduce to
their classical value, and therefore the well-known classical results
for a Maxwellian plasma \cite{landau} are recovered, as it should be.

In \Section{sec:expparam} we define the exponential parametrization
of the nonextensive distribution.
There we also define the analog of the so-called marginal distribution
that enters in the calculation of the dielectric permittivity, and we
discuss some subtleties that must be kept in mind when using the concept
of the marginal distribution in the nonextensive case.
In \Section{sec:genlongdielectric}, we consider the calculation of
the (longitudinal) dielectric permittivity in various cases that we have
mentioned. We first consider the case of the ultrarelativistic plasma, and then
the non-relativistic plasma in both
the low and high frequency limits.
Finally \Section{sec:conclusions} contains our conclusions.

\section{Exponential parametrization of the nonextensive distribution}
\label{sec:expparam}

Up to a normalization factor $C_q$, to be determined below,
we write the nonextensive momentum distribution function in the form
\beq
\label{fqdef1}
f_q(\vec p) = C_q \frac{1}{\Gamma(\frac{1}{1-q})} \int_0^\infty dt\;
t^{\left(\frac{1}{1 - q} - 1\right)}
e^{-[1 + \left(1 - q\right){\beta E_p} ]t}\,,
\eeq
where $E_p$ is the kinetic energy
of a particle with momentum $\vec p$
and $1/\beta = k_B T$, with $T$ being the temperature of the system.
In writing \Eq{fqdef1} we have made use of the following
identity for the Gamma function,
\beq
\label{Gammadef}
A^{-\ell} = \frac {1}{\Gamma (\ell)} \int_0^\infty dt\;t^{\ell-1} e^{-At}\,,
\eeq
which is valid is valid for any $A > 0$, and
for $A = 1$ it becomes just the integral formula for the
$\Gamma$ function. In \Eq{fqdef1} we have used it with the
identification $A = (1+(1-q)\beta E_p)$
and $\ell = \frac {1}{1-q}$.
This parametrization is valid in the range $0 < q < 1$.
As we show below, the lower limit is
further restricted by the requirement that the distribution function
can be normalized.

The working assumption involved is that in the calculations
of the quantities of interest we can interchange the order
of the integration over $t$ with the integration over the
momentum variables of the particle number distributions.
It must be kept in mind that, if the limit $q\rightarrow 1$ is
desired, it must be taken after performing the integration over $t$.
The factor $C_q$ is determined by the the normalization condition
\beq
\label{normcond}
n = \int\frac{d ^{3} p}{(2\pi)^{3}} f_{q}(\vec p)\,,
\eeq
where $n$ is the number density of the electrons.
In the general case, for a given momentum $\vec p$,
the kinetic energy $E_p$ 
in \Eq{fqdef1} is given by
\beq
E_p = \sqrt{m^2 + \vec p^{\,2}} - m\,,
\eeq
where $m$ is the electron mass\cite{footnote2}.
Here we consider, separately, either the non-relativistic(NR) or
the extremely-relativistic(ER) limit,
\beq
\label{kineticenergy}
E_p = \left\{
\begin{array}{ll}
  \frac{p^{2}}{2m} & \mbox {(NR)}\\[12pt]
  p  & \mbox {(ER)}
\end{array}
\right.
\eeq
where
\beq
p \equiv |\vec p|\,.
\eeq

\subsection{NR case}

In this case, \Eq{normcond} implies that $C_q$ is given by
\beq
\frac{1}{C_q} = \frac{1}{n} \frac{1}{\Gamma(\frac{1}{1-q})}
\int_0^\infty dt\;
t^{\left(\frac{1}{1-q}-1\right)}
\int\frac{d^{3} p}{(2\pi)^{3}}\;
e^{-[1+(1-q){\beta}\frac {p^{2}}{2m}]t}\,.
\eeq
Performing the Gaussian integration with respect to each momentum coordinate,
\beq
\frac{1}{C_q} = \frac{1}{n}\left(\frac{m}{2\pi\beta}\right)^{3/2}
 \frac{1}{\Gamma(\frac{1}{1-q})(1-q)^{3/2}}
\int_0^\infty dt\; t^{\left(\frac {1}{1-q} - \frac{5}{2}\right)} e^{-t}\,,
\eeq
and finally performing the integration over $t$ leads to
\beq
%\label{eq:norm3d}
C_{q} = n\left(\frac{2\pi\beta}{m}\right)^{3/2}
 \frac{\Gamma(\frac{1}{1-q})(1-q)^{3/2}}
{\Gamma\left(\frac{1}{1-q} - \frac{3}{2}\right)}\,.
\eeq
In summary, in the non-relativistic case,
the exponential parametrization of the nonextensive distribution function,
properly normalized, is given by
\beq
\label{fqnr}
f_q(\vec p) = n\left(\frac{2 \pi \beta}{m}\right)^{3/2}
\frac{(1-q)^{3/2}}{\Gamma\left(\frac{1}{1-q} - \frac{3}{2}\right)}
\int_0^\infty dt\;
t^{\left(\frac {1}{1-q} - 1\right)} e^{-[1 + (1 - q){\beta\frac{p^2}{2m}}]t}\,.
\eeq

In the context of the calculations of the dielectric permittivity,
it is customary to decompose $\vec p$ into its components
parallel ($p_{\parallel}$) and perpendicular ($\vec p_\perp$) to
the wave vector $\vec k$ of the electromagnetic wave, 
since the integral formula for the dielectric
permittivity depends only on $p_\parallel$. It is then
convenient to introduce the \emph{marginal} distribution
[see, for example, \Ref{chenli}, Eq.(18)]
\beq
\label{fqmarginal}
f^{(\parallel)}_q(p_\parallel) \equiv
\int\frac{d^2p_\perp}{(2\pi)^2} f_q(\vec p)\,,
\eeq
in analogy with the procedure in
the standard case[see, for example, Ref.\cite{landau}, p.123].

Using \Eq{fqnr} and performing the
Gaussian integrals over $\vec p_\perp$, we obtain
the following parametrization for the marginal distribution in the
nonextensive case 
\beq
\label{fqmarginalparam}
f^{(\parallel)}_q(p_\parallel) = n\left(\frac{2 \pi \beta}{m}\right)^{1/2}
\frac{(1-q)^{1/2}}{\Gamma\left(\frac{1}{1-q} - \frac{3}{2}\right)}
\int^\infty_0 dt\; F_q(p_\parallel,t)\,,
\eeq
where
\beq
\label{Fq}
F_q(p_\parallel,t) \equiv
t^{\left(\frac {1}{1-q}-2\right)}
e^{-[1 + (1 - q)\beta p_{\parallel}^{2}/2m]t}\,.
\eeq
\Eq{fqmarginalparam} is the formula that we will use in the calculations.
By construction, i.e., \Eq{fqmarginal}, it satisfies
the normalization condition 
\beq
\label{fqmarginalnorm}
\int^\infty_{-\infty}
\frac{dp_\parallel}{2\pi} f^{(\parallel)}_q(p_\parallel) = n\,.
\eeq

It is appropriate to comment at this point the following.
It is not correct to construct $f^{(\parallel)}_q(p_\parallel)$
by taking it to be the generalization of the standard marginal distribution,
i.e., take $f^{(\parallel)}_q(p_\parallel)$ to be
\beq
f^{(\parallel)}_q(p_\parallel) \propto
e_q\left(-\beta p^2_\parallel/2m\right)\,.
\eeq
That would give an expression of the form
\beq
\left(f^{(\parallel)}_q(\vec p)\right)_{incorrect} =
C^\prime_q \frac {1}{\Gamma(\frac{1}{1-q})} \int_0^\infty dt\;
t^{\left(\frac {1}{1-q}-1\right)}
e^{-[1+(1-q){\beta p^{\,2}_\parallel/2m} ]t}\,,
\eeq
which is incorrect even if $C^\prime_q$ is chosen such that
it is normalized according to \Eq{fqmarginalnorm}.
This is ultimately a consequence of the fact that the
generalized exponential exponential function $e_q(A)$ does not
satisfy the relation
\beq
\label{exponentialextensiveprop}
e(A + B) = e(A) e(B)\,.
\eeq
that the exponential function does.
Therefore, a calculation in the nonextensive case,
based on a marginal distribution assumed to be the generalization
of the standard classical marginal distribution leads to incorrect results.

\subsection{ER case}

In this case the normalization condition implies that $C_q$ is given by
\beq
\frac{1}{C_q} = \frac{1}{n}\frac{1}{\Gamma(\frac{1}{1-q})}
\int_{0}^{\infty} t^{\frac{1}{1-q}-1} e^{-t}
dt \int \frac{d^{3}p}{(2\pi)^3}
e^{-(1-q)\beta pt}\,.
\eeq
Straightforward evaluation of both integrals then yields
\beq
C_{q} = \frac{n\beta^{3}}{8\pi}(1-q)^{3}
\frac{\Gamma(\frac{1}{1-q})}{\Gamma(\frac{1}{1-q} - 3)}\,,
\eeq
and therefore
\beq
\label{fqer}
f_{q}(p) = \frac{n\beta^{3}}{8\pi}
\frac{(1-q)^{3}}{\Gamma(\frac{1}{1-q} - 3)}
\int_{0}^{\infty}dt\; t^{\frac{1}{1-q}-1} e^{-[1+(1-q)\beta p]t}\,.
\eeq

\subsection{Valid range of $q$}

  As already mentioned above, although the exponential parametrization is
  valid for $q$ in the range $0 < q < 1$, the lower limit is
  further restricted by the requirement that the distribution function
  can be normalized. Since, for real $x$, $\Gamma(x)$ is defined provided
  $x > 0$, looking at \Eqs{fqnr}{fqer}  we see that $q$ must be restricted
  such that
\beq
\label{qcondnr}
\frac{1}{1 - q} - \frac{3}{2} > 0\quad \mbox{(NR case)}
\eeq
or
\beq
\label{qconder}
\frac{1}{1 - q} - 3 > 0\quad \mbox{(ER case)}
\eeq
otherwise the distribution functions have non-physical
value or, stated diferently, the distribution function in \Eq{fqdef1}  cannot
be normalized. Equations (\ref{qcondnr}) and (\ref{qconder}) yield
the restrictions that we adopt throughout,
\beq
q > \frac{1}{3} \quad \mbox{(NR case)}
\eeq
or
\beq
q > \frac{2}{3} \quad \mbox{(ER case)}
\eeq

\section{Longitudinal dielectric permittivity}
\label{sec:genlongdielectric}

We now consider the application of the exponential parametrization
of the nonextensive distribution to the calculation
of the longitudinal dielectric permittivity and dispersion relation.
The starting point is the standard formula for the
longitudinal dielectric permittivity for a collisionless electron
plasma for a wave with wave vector $\vec k$ and frequency
$\omega$\cite{landau,nichol},
\beq
\label{epsilonlong}
\epsilon_{l} = 1 - \frac {4 \pi e^2}{k^2}
\int\frac{d^{3}p}{(2 \pi)^{3}}
\frac{\vec k\cdot \nabla_{\vec p}f_q(\vec p)}
{(\vec k \cdot \vec v - \omega -i \epsilon)}\,,
\eeq
where $\vec v = \vec p/m$ is the electron 
  velocity. As usual, the formula includes the factor $i\epsilon$
  for defining the function at the pole $\omega = \vec k \cdot \vec v$
  according to the Landau rule. We consider the evaluation of
  \Eq{epsilonlong}, separately depending on whether the gas
  is non-relativistic or ultrarelativistic.

\subsection{Ultra-relativistic (ER) case}
\label{sec:relativistic}

In this case we set $v \rightarrow 1$ (remembering
  our convention of using natural units throughout, as we already
  stated\cite{footnote2}).
Denoting by $\theta$ the angle between $\vec p$ and $\vec k$,
in this case we then have
\beq
\vec k\cdot\vec v = k\cos\theta\,,
\eeq
where
\beq
k = |\vec k|\,,
\eeq
and
\beq
\vec k\cdot \vec \nabla_{p} f_q = k\cos\theta\frac{\partial f_q}{\partial p}\,,
\eeq
where, from \Eq{fqer},
\beq
\frac{\partial f_{q}}{\partial p} =
- \frac{n\beta^{4}}{8\pi} \frac{(1-q)^{4}}{\Gamma(\frac{1}{1-q} -3)}
\int_{0}^{\infty}t^{\frac{1}{1-q}}e^{-[1+(1-q)\beta p]t} dt\,.
\eeq
Therefore, from \Eq{epsilonlong}
\beqa
\label{eq:dielectricultrarelativistic}
\epsilon_{l} & = & 1 - \frac{8\pi^2e^{2}}{k}
\int dp\,p^2 d(\cos\theta)\frac{\frac{\partial f_{q}(p)}{\partial p}
  \cos{\theta}}{k\cos{\theta} - \omega -i\epsilon}\nonumber\\
& = & 1 + \frac{n\pi e^{2} \beta^{4}}{k}
\frac{(1-q)^{4}}{\Gamma(\frac{1}{1-q} -3)}
\left[
\int_{-1}^{1}d\cos\theta\frac{\cos{\theta}}{k\cos{\theta} - \omega -i\epsilon}
\right] \times
\nonumber\\
&&
\int_{0}^{\infty}dt\, t^{\frac{1}{1-q}}e^{-t}
\int_{0}^{\infty} dp\,p^2 e^{-(1-q)\beta pt}\,.
\eeqa
The integral over $p$ is
\beq
\int_{0}^{\infty}dp\, p^2 e^{-(1-q)\beta pt}  =
\frac{2}{(1-q)^{3}\beta^{3} t^{3}}\,,
\eeq
and for the resulting integral over $t$ we use
\beq
\int_{0}^{\infty}dt\, t^{\frac{1}{1-q}-3}e^{-t} =
\Gamma\left(\frac{1}{1-q}-2\right)\,.
\eeq
Finally, the integral over the angle $\theta$ is evaluated by rewriting it
in the form
\beq
\frac{1}{k} \int^{1}_{-1} dx \frac{x}{x-\frac{\omega}{k}-i\epsilon}\,,
\eeq
and using the identity(see \cite{landau})
\beq
\label{princvalue}
\frac{1}{(x-a) \mp i \epsilon} = P\frac{1}{x-a} \pm i \pi \delta(x-a)\,,
\eeq
where $\delta(x-a)$ is the Dirac delta function and $P$ denotes
the principal value. Thus we obtain 
\beq
\label{epsilonrerlimit}
\epsilon_{\ell} = 1 + \frac{4\pi e^2 n}{k^2 T_{e}}(3q-2)
\left\{
  1 + \frac{\omega}{2k}\log\left|\frac{\omega - k}{\omega + k}\right|
+ i \frac{\pi \omega}{2k}\right\}\,,
\eeq
where we have put
\beq
T_e \equiv \frac{1}{\beta} = k_B T\,.
\eeq

As anticipated in \Section{sec:introduction}, the result given in
\Eq{epsilonrerlimit} coincides with the result obtained previously in
\Ref{liu} with the commonly-used parametrization
  of the non-extensive distribution.
\Eq{epsilonrerlimit} corresponds to the
  formula obtained by making the replacement
\beq
\label{Tqer}
T\rightarrow \frac{T}{3q - 2}\,,
\eeq
in the well-known result for the classical Maxwell-Boltzman
case (see, for example \Ref{landau}, pp. 132).
By the same token, the dispersion relations in the
  present case, defined by the condition $\epsilon_{\ell} = 0$,
  can be obtained from the corresponding dispersion relations
  of the classical Maxwell-Boltzman case by making the same replacement
  given in \Eq{Tqer}. For example, in the limit $\omega \gg k$,
\begin{equation}
\omega_{k}^{2} = (\Omega^{(q)}_{e,rel})^{2} + \frac{3}{5}k^{2}
\end{equation}
where
\beq
(\Omega^{(q)}_{e,rel})^{2}  \equiv \frac{4\pi e^2 n}{3 T_{e}}(3q-2)\,.
\eeq
Evidently, \Eq{epsilonrerlimit} and the dispersion relations
reduce to the results for the classical Maxwell-Boltzman case
in the limit $q \rightarrow 1$.

As already mentioned above,
  the result given in \Eq{epsilonrerlimit} coincides with
  the result obtained previously in \Ref{liu}.
  The ability to reproduce the result, coupled with the ease
  with which it has been obtained here, gives us
  confidence in the capability of the exponential parametrization
  method for this kind of calculation.

\subsection{Non-relativistic (NR) case}
\label{subsecsec:nonrelativistic}

In this case a useful expression for
  the dielectric constant in terms of an integral over the marginal
  distribution can be obtained as follows. Using \Eq{fqmarginal},
the formula for $\epsilon_{l}$ given in \Eq{epsilonlong} becomes
\beq
\epsilon_{l} = 1 - \frac {4 \pi e^2}{k^2}
\int^\infty_{-\infty}\frac{dp_\parallel}{2\pi}
\frac{k\frac{\partial f^{(\parallel)}_q}{\partial p_\parallel}}
{\frac{kp_\parallel}{m} - \omega -i \epsilon}\,,
\eeq
where we have used
\beq
\vec k\cdot\vec p = kp_\parallel\,,
\eeq
Using \Eqs{fqmarginalparam}{Fq}, together with
\beq
\frac{\partial F_q}{\partial p_\parallel} = 
-\frac{(1-q)\beta t p_\parallel}{m}F_q\,,
\eeq
it follows that
\beq
\label{epsilonIrel}
\epsilon_{l} = 1 + \frac {4 \pi e^2 n\delta}{k^2}\int^\infty_0 dt\;
t^{\left(\frac{1}{1-q} - 1\right)} e^{-t}I\,,
\eeq
where we have defined
\beq
\label{Idef}
I \equiv (2\pi)^{1/2} k\int\frac{dp_\parallel}{2\pi} 
\frac{p_\parallel e^{-(1-q){\beta}
\frac{p^2_\parallel}{2m}t}}{\frac{kp_\parallel}{m} - \omega -i \epsilon}\,,
\eeq
and
\beq
\label{eq:delta}
\delta = \left(\frac{\beta }{m}\right)^{3/2}
\frac{(1-q)^{3/2}}{\Gamma\left(\frac{1}{1-q} - \frac{3}{2}\right)}\,.
\eeq
Using the the identity \Eq{princvalue} once again, we obtain
\beq
\label{eq:I}
I = I^{(r)} + i\sqrt{\frac{\pi}{2}}\frac{m^2\omega}{k}
e^{-(1-q)\beta\frac{m\omega^{2}}{2k^{2}}t} \,,
\eeq
with
\beq
\label{Isubr}
I^{(r)} = (2\pi)^{1/2} k P\int^\infty_{-\infty}\frac{dp_\parallel}{2\pi} 
\frac{p_\parallel e^{-(1-q){\beta}
\frac{p^2_\parallel}{2m}t}}{\frac{kp_\parallel}{m} - \omega}\,,
\eeq

In what follows we focus exclusively on the real part of
the dielectric permittivity, which we denote by $\epsilon^{(r)}_{l}$.
From \Eqss{epsilonIrel}{eq:I}{Isubr}, it is given by
\beq
\label{basicintegralformula}
\epsilon^{(r)}_{\ell} = 1 + \frac {4 \pi e^2 n\delta}{k^2}\int^\infty_0 dt\;
t^{\left(\frac{1}{1-q} - 1\right)} e^{-t}I^{(r)}\,,
\eeq
with $\delta$ given in \Eq{eq:delta}.

\Eq{basicintegralformula} is our basic integral formula using
the exponential parametrization of the nonextensive distribution
in the NR case, which can be evaluated
explicitly by considering special cases or limits of interest.
We consider next the explicit evaluation in the high or
low frequency limit.

\subsubsection{NR High frequency limit}
\label{sec:nrhighfreqlimit}

In this case, $\omega \gg kp_\parallel/m$.
We rewrite \Eq{Isubr} in the form
\beq
I^{(r)} = -(2\pi)^{1/2}\left(\frac{k}{\omega}\right)
P\int^\infty_{-\infty}\frac{dp_\parallel}{2\pi} 
\frac{p_\parallel e^{-(1-q){\beta}\frac{p^2_\parallel}{2m}t}}
{1 - \frac{kp_\parallel}{m\omega}}\,.
\eeq
The result obtained in \Ref{chenli} for this case corresponds to assume
that $\omega,k$ and the parameters of the distribution function are such that 
the denominator can be expanded in powers of
$\frac{kp_\parallel}{m\omega}$ and then evaluating the resulting integrals term
by term. Thus,
%
%\beq
%\begin{split}
%(1 - k p_{z} / m\omega)^{-1} &= 1 + k p_{z} / m\omega +
%$\frac{(-1)(-2)}{2!} (k %p_{z} / m\omega)^{2} \\
%&+ \frac{(-1)(-2)(-3)}{3!}(-k p_{z} / m\omega)^{3}+....
%\end{split}
%\end{equation}
%
\beq
I^{(r)} = -(2\pi)^{1/2}\left(\frac{k}{\omega}\right)^2\frac{1}{m}
\int^\infty_{-\infty}
\frac{dp_\parallel}{2\pi}\,e^{-(1-q){\beta}\frac{p^2_\parallel}{2m}t}
\left[p^2_\parallel + \left(\frac{k}{m\omega}\right)^2 p^4_\parallel\right]\,,
\eeq
where the odd terms in $p_\parallel$ have been dropped since they integrate
to zero. Performing the Gaussian integration of the remaining ones then yield
%
%\begin{equation}
%	\begin{split}
%\Re[I] &= \xi  (-k/m \omega) \wp \int p_{z} e^{-(1-q){\beta}
% p_{z}^{2}t/2m}[1 %+ k p_{z} / m\omega +  (k p_{z} / m\omega)^{2} \\
%&+ (k p_{z} / m\omega)^{3})]  \frac{1}{(2 \pi)^{3}} \ dp_{z}
%	\end{split}
%\end{equation}
%
\beqa
I^{(r)} & = & -\left(\frac{m}{\beta}\right)^{\frac{3}{2}}
\frac{k^2}{m\omega^2}\left\{
\frac{t^{-\frac{3}{2}}}{(1 - q)^{\frac{3}{2}}} +
\frac{3}{\beta m}\frac{k^2}{\omega^2}
\frac{t^{-\frac{5}{2}}}{(1 - q)^{\frac{5}{2}}}\right\}\,.
\eeqa
Substituting this result in \Eq{basicintegralformula} and using
the definition of the Gamma function
[i.e., \Eq{Gammadef} with $A = 1$ 
  and $\ell = \frac{1}{1 - q} - \frac{3}{2}$ or
  $\frac{1}{1 - q} - \frac{5}{2}$]
then yields
\beq
\epsilon^{(r)}_\ell = 1 - \frac{\omega^2_p}{\omega^2}\delta
\left(\frac{m}{\beta}\right)^{\frac{3}{2}}\left[
\frac{\Gamma\left(\frac{1}{1 - q} - \frac{3}{2}\right)}{(1 - q)^{\frac{3}{2}}}
+
\frac{3}{\beta m}\frac{k^2}{\omega^2}
\frac{\Gamma\left(\frac{1}{1 - q} - \frac{5}{2}\right)}{(1 - q)^{\frac{5}{2}}}
\right]\,,
\eeq
where $\omega_p$ is the plasma frequency,
\beq
\label{omegap}
\omega^2_p = \frac{4\pi^2 n}{m}.
\eeq
Using \Eq{eq:delta} and remembering that $\Gamma(z) = \frac{1}{z}\Gamma(z+1)$
[putting in our case $z = \left(\frac{1}{1-q} - \frac{5}{2}\right)$]
we then obtain,
\beq
\label{epsilonhighfreqlimitfinal}
\epsilon^{(r)}_\ell = 1 - 
\frac{\omega_{p}^{2}}{\omega^{2}}
\left[1 + \frac{3k^2 v^2_T}{\omega^2}\frac{2}{5q-3}\right]\,,
\eeq
with
\beq
\label{vT}
v_T \equiv \sqrt{\frac{1}{m\beta}}\,.
\eeq
The corresponding dispersion relation
obtained by the condition $\epsilon^{(r)}_\ell = 0$ is then
\beq
\label{disprelhighfreqlimitfinal}
\omega^{2}_k = \omega_{p}^{2} + 3 k^{2}v_{T}^{2}\frac{2}{5q-3}
\eeq
The results in \Eqs{epsilonhighfreqlimitfinal}{disprelhighfreqlimitfinal}
coincide with those obtained by Chen and Li\cite{chenli}.
In the limit $q = 1$, they reduce to the classical formulas.
By making contact with a known result, our purpose here has been
to show and convince ourselves once more that the exponential parametrization
of the nonextensive distribution
provides a succinct method for carrying out this type of calculation.
To assess the effectiveness of the exponential parametrization method,
the reader is invited to consult the corresponding calculation in \Ref{chenli}
of the result given in Eq. (38) of that reference.

\subsubsection{NR Low frequency limit}
\label{sec:nrlowfreqlimit}

We now consider the low frequency limit, that is
$\omega \ll \frac{kp_\parallel}{m}$. As mentioned
in the Introduction, the formula for this limit case,
can be deduced from the calculations of \Ref{dai}, who consider the
stream instability induced by ions and electrons with
different drift velocities in a dusty plasma,
by adapting and restricting the attention to the electron contribution.
Our own result confirms the result obtained that way.
With the exponential parametrization method
the formula for this case also follows straightforwardly,
and as a byproduct we show that the next-to-leading order contribution
to the dielectric permittivity (in powers of $\omega/k$) is determined
equally simply, which is not considered in \Ref{dai}.
  
We rewrite \Eq{Isubr} in the form
\beq
\label{Isubrlowfreq}
I^{(r)} = m P\int^\infty_{-\infty}\frac{dp_\parallel}{\sqrt{2\pi}} 
\frac{p_\parallel e^{-(1-q){\beta}
\frac{p^2_\parallel}{2m}t}}{p_\parallel - \frac{m\omega}{k}}\,.
\eeq
Writing
$p_\parallel = (p_\parallel - \frac{m\omega}{k}) + \frac{m\omega}{k}$
in the numerator,
\beq
\label{defIsubrlowfreq1}
I^{(r)} = m^{3/2}\left(\frac{1}{(1 - q)\beta t}\right)^{1/2} + I^{(r)}_1\,,
\eeq
where
\beq
I^{(r)}_1 = \frac{m^2\omega}{k} P\int^\infty_{-\infty}
\frac{dp_\parallel}{\sqrt{2\pi}} 
\frac{e^{-(1-q){\beta}\frac{p^2_\parallel}{2m}t}}
{p_\parallel - \frac{m\omega}{k}}\,,
\eeq
and by the change of variable $p_\parallel = \frac{m\omega}{k} + u$,
\beq
\label{Ir}
I^{(r)}_1 = \frac{m^2\omega}{k}
e^{-(1 - q)\frac{\beta m\omega^2}{2k^2}t}
P\int^\infty_{-\infty}\frac{du}{\sqrt{2\pi}}\frac{1}{u}
e^{-(1 - q)\frac{\beta u^2}{2m}t} e^{-(1 - q)\frac{\beta\omega u}{k}t}\,.
\eeq
Expanding the last exponential factor in \Eq{Ir}
up to the linear term in $u$,
\beq
e^{-(1 - q)\frac{\beta\omega u}{k}t} \simeq 1 -
(1 - q)\frac{\beta\omega u}{k}t\,,
\eeq
the $1/u$ term is odd and integrates to zero while the remaining integral
is a Gaussian. Thus,
\beqa
I^{(r)}_1 & = & -(1 - q)\frac{\beta m^2\omega^2 t}{k^2}
e^{-(1 - q)\frac{\beta m\omega^2}{2k^2}t}
\sqrt{\frac{m}{(1 - q)\beta t}}\nonumber\\
& = &
-\frac{m^{5/2}\omega^2}{k^2}
\left[(1 - q)\beta t\right]^{1/2}
e^{-(1 - q)\frac{\beta m\omega^2}{2k^2}t}\nonumber\\
& = &
-\frac{m^{5/2}\omega^2}{k^2}
\left[(1 - q)\beta t\right]^{1/2} + O\left(\frac{\omega^4}{k^4}\right)\,,
\eeqa
and using this result in \Eq{defIsubrlowfreq1}, we then have
\beq
\label{Isubrlowfreqfinal}
I^{(r)} = m^{3/2}\left(\frac{1}{(1 - q)\beta t}\right)^{1/2}\left[1 -
\frac{m\omega^2}{k^2}(1 - q)\beta t\right]\,,
\eeq
dropping the terms $O\left(\frac{\omega^4}{k^4}\right)$.
Substituting \Eq{Isubrlowfreqfinal} in \Eq{basicintegralformula},
\beqa
\epsilon^{(r)}_\ell & = & 1 + \frac{4\pi e^2 n}{k^2}
\frac{\beta(1 - q)}{\Gamma\left(\frac{1}{1 - q} - \frac{3}{2}\right)}
\int^\infty_0 dt\; t^{\left(\frac{1}{1 - q} - \frac{3}{2}\right)} e^{-t}
\left[1 - (1 - q)\frac{m\omega^2\beta}{k^2}t\right]\nonumber\\
& = & 1 + \frac{4\pi e^2 n}{k^2}
\frac{\beta(1 - q)}{\Gamma\left(\frac{1}{1 - q} - \frac{3}{2}\right)}\times
\nonumber\\
&&
\left[\Gamma\left(\frac{1}{1 - q} - \frac{1}{2}\right) -
(1 - q)\frac{m\omega^2\beta}{k^2}
\Gamma\left(\frac{1}{1 - q} + \frac{1}{2}\right)\right]\,.
\eeqa

Using once again the property $\Gamma(z + 1) = z\Gamma(z)$, after
some straightforward algebra this simplifies to
\beq
\label{epsilonrlowfreqfinal}
\epsilon^{(r)}_\ell = 1 + \frac{\omega^2_p}{v^2_T k^2}
\left(\frac{3q - 1}{2}\right)
\left[1 - \frac{\omega^2}{v^2_T k^2}\left(\frac{1 + q}{2}\right)\right]\,,
\eeq
with $\omega_o$ and $v_T$ defined in \Eqs{omegap}{vT}, respectively.
The corresponding dispersion relation is then given by
\beq
\label{disprellowfreqfinal}
\omega^{2}(k) = \left(\frac{2}{q+1}\right) k^2 v^2_T
\left[1 + k^{2}\lambda^2_D \left(\frac{2}{3q-1}\right)\right]\,,
\eeq
where
\beq
\lambda_D \equiv \frac{v_T}{\omega_p}\,,
\eeq
is the Debye wavelength. As expected, the results given in
\Eqs{epsilonrlowfreqfinal}{disprellowfreqfinal}
reduce to the classical formulas
in the limit $q \rightarrow 1$ (see e.g., \Ref{landau}, pp. 137),

  As already pointed out above, the exponential parametrization method
  allows us to reproduce the results obtained previously in \Ref{dai},
  with much less effort, and as further indication of this, we have
  obtained the next-to-leading term in Eq.(\ref{epsilonrlowfreqfinal}).

\section{Conclusions and Outlook}
\label{sec:conclusions}

In this article we have introduced and illustrated the use the exponential
parametrization method of the nonextensive distribution
in the calculations of the dielectric permittivity and the dispersion
relation of a collisionless electron plasma. We have established contact
with some of the results that have been obtained previously by other other
authors by other means in various situations or limiting cases.
The results we have obtained illustrate the effectiveness and simplicity
of the method, and paves the way for considering
the application of the method to other cases of interest,
in the context of similar or related systems, such as magnetized
and/or anisotropic plasmas.

\section*{Author contribution statement}

The authors of this paper have contributed equally to all the aspects of this work and its presentation.


\begin{thebibliography}{99}

\bibitem{tsallis} C. Tsallis,
  \emph{Possible Generalization of Boltzmann-Gibbs Statistics},
  Stat Phys 52, 479 (1988). https://doi.org/10.1007/BF01016429.

\bibitem{tsallis2} C. Tsallis,  F. Baldovin,  R. Cerbino and P. Pierobon,
  \emph{Introduction to Nonextensive Statistical Mechanics and Thermodynamics},
  (2003) [arXiv:cond-mat/0309093v1 [cond-mat.stat-mech]].

\bibitem{oikonomou} T. Oikonomou, G.B. Bagci,
  Phys. Lett. A 374, 2225 (2010);
  Phys. Lett. A 381, 207 (2017);
  Phys. Rev. E 97, 012104 (2018);
  Phys. Rev. E 99, 032134 (2019);
  
\bibitem{presse} S. Presse et al,
  Phys. Rev. Lett. 111, 180604 (2013)

\bibitem{bidollina} Aruna Bidollina, Thomas Oikonomou, G. Baris Bagci,
  \emph{Opening Pandora’s Box: Maximizing the $q$-entropy with
  Escort Averages}, [arXiv:1904.00581 [cond-mat.stat-mech]].

\bibitem{footnote1} The use of nonextensive statistics raises many questions
related to thermodynamic anomalies and other possible fundamental
inconsistencies, several of which are discussed in the referencences
cited in \cite{oikonomou,presse,bidollina}. Leaving those issues aside,
our purpose is to point out that some specific calculations that
have been considered with them in the literature are greatly simplified
by using the method we present.

\bibitem{dash} Sadhana Dash and D. P. Mahapatra,
\emph{Transverse Momentum Distribution in Heavy Ion Collision using
$q$-Weibull Formalism}, (2016)
[arXiv:1611.04025v2 [nucl-th]].

\bibitem{smbat} Smbat Grigoryan,
\emph{Using the Tsallis distribution for hadron spectra in $pp$ collisions:
  Pions and quarkonia at $\sqrt{s}$ = $5-13000$ GeV.},
Phys. Rev. D 95, 056021 (2017) [arXiv:1702.04110v4 [hep-ph]].

\bibitem{lima} J.A.S. Lima, R. Silva, Jr., and Janilo Santos,
  \emph{Plasma oscillations and nonextensive statistics},
  Phys. Rev. E 61, (2000).

\bibitem{chenli} X. C. Chen and  X. Q. Li,
  \emph{Comment on Plasma oscillations and nonextensive statistics},
  Phys. Rev. E 86, 068401 (2012) [arXiv:1206.2345v1 [physics.plasm-ph]].

\bibitem{liyan} Liu Liyan and Du Jiulin,
  \emph{Ion acoustic waves in the plasma with the power-law -distribution
    in nonextensive statistics},
  Physica A 387, 4821 (2008),
  [arXiv:0804.3732v1 [physics.plasm-ph]].

%\bibitem{zhipeng} Liu Zhipeng, Du Jiulin and Guo Lina,
%  \emph{Nonextensivity and q-distribution of a relativistic gas under
%    an external electromagnetic field}, 
%  Chin. Sci. Bull. 56, 3689–3692 (2011),
%  [arXiv:0802.2492v1 [cond-mat.stat-mech]].

\bibitem{munoz} V\'ictor Mu\~noz,
  \emph{Longitudinal Oscillations in a Nonextensive Relativistic Plasma}, 
  Nonlin. Processes Geophys 13, 237 (2006),
  [arXiv:physics/0410204v1 [physics.plasm-ph]].

\bibitem{liu} San-Qiu Liu, Xiao-Chang Chen,
  \emph{Dispersion relation of longitudinal oscillation in
    relativistic plasmas with nonextensive distribution},
  Physica A 390, 1704 (2011).

\bibitem{dai} Jin-Wei Dai, Xiao-Chang Chen and Xiao-Qing Li,
  \emph{Dust ion acoustic instability with q-distribution in nonextensive
    statistics},
    Astrophys. Space Sci. 346, 183 (2013).

\bibitem{pain} J. C. Pain,  D. Teychenn\'e and F. Gilleron,
  \emph{Self-consistent modelling of hot plasmas within non-extensive
    Tsallis' thermostatistics},
  Eur. Phys. J. D 65, 441–445 (2011)
  [arxiv.org:1110.0446v1]

\bibitem{footnote2} We use natural units throughout,
  and therefore we set $c = 1$ from now on.

\bibitem{landau} E. M. Lifshitz and L.P. Pitaevskii,
  \emph{Physical Kinetics: Landau and Lifshitz Course of Theoretical Physics}, 
  Volume 10, Pergamon Press Ltd; 1st Edition, (1981) p. 121-124, 128-132.

\bibitem{nichol} Dwight R. Nicholson, \emph{Introduction to Plasma Theory},
  John Wiley and Sons, (1983), p. 70-76.

\bibitem{krall} Nicholas A. Krall and Alvin W. Trivelpiece,
  \emph{Principles of Plasma Physics}, McGraw-Hill, Inc. (1973), p. 381-389
\end{thebibliography}
\end{document}